\documentclass[doublecol]{epl2} 

\usepackage{nameref}
\usepackage{amsmath, amssymb, latexsym, MnSymbol, bm}
\usepackage{blkarray}
\usepackage[table]{xcolor}
\usepackage{xcolor}
\usepackage[final]{pdfpages}

\usepackage[%
  colorlinks=true,
  urlcolor=blue,
  linkcolor=blue,
  citecolor=blue
]{hyperref}

\title{Clustering free-falling paper motion with complexity and entropy}

\author{
Arthur A. B. Pessa\inst{1,\hspace*{-0.5em}}\thanks{\email{arthur\_pessa@hotmail.com}} 
\and 
Matja{\v z} Perc\inst{2,3,4,5,\hspace*{-0.5em}}\thanks{\email{matjaz.perc@gmail.com}} 
\and 
Haroldo V. Ribeiro\inst{1,\hspace*{-0.5em}}\thanks{\email{hvr@dfi.uem.br}} 
}

\shortauthor{Pessa, Perc, and Ribeiro}

\institute{                    
  \inst{1} Departamento de F\'isica, Universidade Estadual de Maring\'a - Maring\'a, PR 87020-900, Brazil\\
  \inst{2} Faculty of Natural Sciences and Mathematics, University of Maribor, Koro{\v s}ka cesta 160, 2000 Maribor, Slovenia\\
  \inst{3} Department of Medical Research, China Medical University Hospital, China Medical University, Taichung, Taiwan\\
  \inst{4} Alma Mater Europaea, Slovenska ulica 17, 2000 Maribor, Slovenia\\
  \inst{5} Complexity Science Hub Vienna, Josefst{\"a}dterstra{\ss}e 39, 1080 Vienna, Austria
}

\abstract{Many simple natural phenomena are characterized by complex motion that appears random at first glance, but that often displays underlying patterns and behavior that can be clustered in groups. The movement of small pieces of paper falling through the air is one of these systems whose complete mathematical description seems unworkable. Understanding these types of motion thus demands automated experimentation capable of producing large datasets covering different behaviors -- a task that has become feasible only recently with advances in computer vision and machine learning methods. Here we use one of these datasets related to the motion of free-falling paper with different shapes to propose an information-theoretical approach that automatically clusters different types of behavior. We evaluate the permutation entropy and statistical complexity from time series related to the observable area of free-falling paper pieces captured by a video camera. We find that chaotic and tumbling motions have a distinct average degree of entropy and complexity, allowing us to accurately discriminate between these two types of behavior with a simple unsupervised machine learning algorithm. Our method has a performance comparable to other approaches based on physical quantities but does not depend on reconstructing the three-dimensional falling trajectory.}

\begin{document}

\maketitle

\section{Introduction}\label{sec:introduction}

The motion of falling objects has fascinated researchers since the works of Aristotle~\cite{dugas1955history}. These investigations were fundamental for the development of classical mechanics and gravitation as well as for physics as a whole~\cite{dugas1955history, rovelli2015aristotle}. Primarily known for its regularities and simple physical description, the motion of falling objects can however become remarkably complex in several everyday situations. That is the case of simple pieces of paper falling through the air, whose complete mathematical description remains a great challenge at present day despite a long history of research on this issue~\cite{tanabe1994behavior, belmonte1998fromflutter, mahadevan1999tumbling}.

Dating back to the seminal work of Maxwell on the tumbling motion of paper pieces~\cite{maxwell1853particular}, the primary approach for addressing this complex motion relies on modeling via fundamental physics~\cite{tanabe1994behavior, belmonte1998fromflutter, andersen2005analysis}.
A notorious example includes the use of a two-dimensional diagram comprising the values of the non-dimensional moment of inertia and the Reynolds number to distinguish among steady, tumbling, and chaotic motions~\cite{willmarth1964steady, field1997chaotic, howison2020large-scale}.
However, recent investigations have shown that the falling-paper problem can be empirically approached through large-scale automated experiments based on computer vision and machine learning methods~\cite{howison2019physics, howison2020large-scale}. This is much in line  with many recent computational developments involving complex systems in areas as diverse as materials science~\cite{butler2018machine, jha2018elemnet, sigaki2019estimating}, traffic management~\cite{li2022analysis, chen2022research}, art history~\cite{sigaki2018history, lee2020dissecting} and science of science\cite{fortunato2018science, weis2021learning}.

The works of Howison and coworkers~\cite{howison2019physics, howison2020large-scale} are pioneering in demonstrating the usefulness of an automated physical experimentation system for a better understanding of the motion of falling paper. Here we rely on their experimental results to investigate and classify time series related to the motion of different free-falling paper shapes. Differently from most previous works, we use an information-theory framework based on ordinal patterns to investigate the falling-paper problem. Our approach avoids the cumbersome mathematical description of the falling-paper motion by calculating two complexity measures --- permutation entropy~\cite{bandt2002permutation} and statistical complexity~\cite{lopezruiz1995statistical, rosso2007distinguishing} --- related to the local ordering patterns obtained from time series associated with the observable area of the pieces of paper captured by video cameras. The combined use of these two complexity measures is usually known as the complexity-entropy plane, an information-theory tool originally proposed for distinguishing between chaotic and stochastic time series~\cite{rosso2007distinguishing} but that has also been successfully used in many other applications~\cite{zunino2010complexity, zunino2012efficiency, ribeiro2012songs, sigaki2019estimating, alves2020collective, ayyad2021cellular, zanin2021ordinal, jara2021using, valensise2021entropy, ma2022complexity}. Beyond representing a simple and direct description, our use of information-theory measures can be further motivated by the strong connection between falling-paper motion and chaos theory~\cite{tanabe1994behavior, field1997chaotic}. 
 
Our results demonstrate that time series associated with tumbling and chaotic motions occupy distinct regions in the complexity-entropy plane, allowing us to accurately discriminate between these two types of behaviors. Indeed, by using an unsupervised learning technique (the $k$-means clustering algorithm~\cite{james2014introduction}), we show that one can automatically distinguish between the two behaviors with accuracy at least comparable to an approach based on estimating the non-dimensional moment of inertia and the Reynolds number~\cite{howison2020large-scale}. Furthermore, our method relies only on one-dimensional time series (observable paper area), while the two physical measures require the complete reconstruction of the three-dimensional trajectory of the center of mass of the paper pieces.

In what follows, we briefly describe the automated experiments of Howison \textit{et al.}~\cite{howison2020large-scale} and the time series used in our work. Next, we present the complexity-entropy plane approach, followed by our main results about the discrimination between the tumbling and chaotic motions of falling papers. Finally, we finish this letter with some concluding remarks.

\section{Data}
All data analyzed in our work were obtained from the article of Howison~\textit{et al.}~\cite{howison2020large-scale} and are freely available at \url{https://github.com/th533/Falling-Paper}. In this reference, the authors combine robotics and computer vision to create an automated system capable of fabricating, dropping, and tracking the free-falling motion of small pieces of paper with different shapes. For the tracking, Howison~\textit{et al.}~\cite{howison2020large-scale} rely on high-speed cameras, which in addition to allowing the three-dimensional reconstruction of the center of mass of the paper pieces, further yield time series of the observable area, that is, the fraction of the paper pieces visible to the cameras as a function of the time. Because the process is fully automated, the authors were able to track and analyze the free-falling motion of hundreds of paper sheets with circular, hexagonal, square, and cross shapes. Furthermore, Howison~\textit{et al.}~\cite{howison2020large-scale} provide a visual classification produced by a panel of human observers (experts) of the free-falling behavior into three types: steady, tumbling, and chaotic. Steady falls are the most straightforward behavior characterized by an approximately straight vertical motion of the center of mass combined with small oscillations along the horizontal direction. Tumbling falls are marked by a continuous end-over-end motion combined with a lateral drift. In contrast, the chaotic falls alternate between tumbling and large swooping motions with no apparent regularity. 

\begin{figure*}[!ht]
\centering
\includegraphics[width=0.95\linewidth]{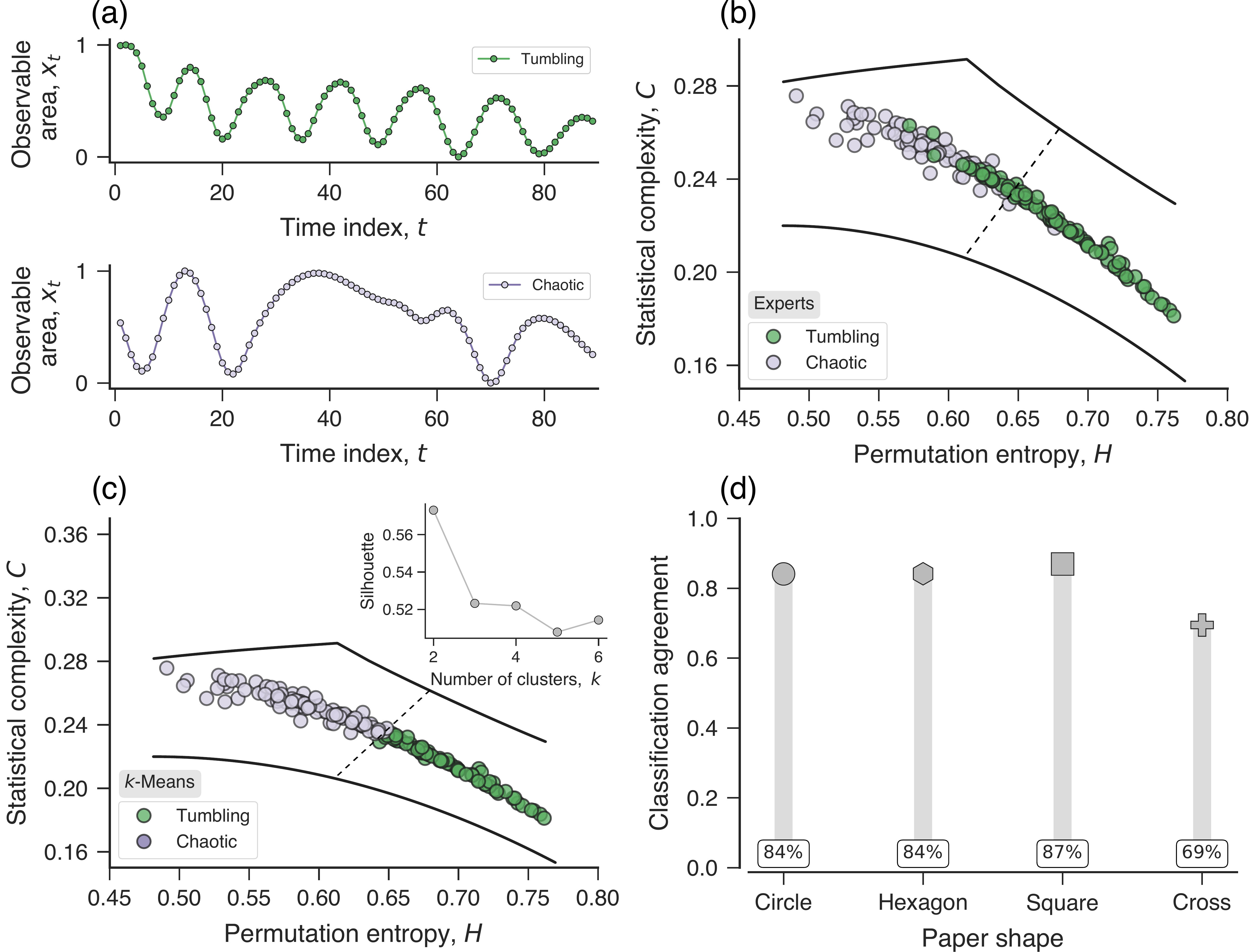}
\caption{Using the complexity-entropy plane to distinguish between tumbling and chaotic free-falling motions of papers sheets. (a) Two typical time series of the observable area of circular paper sheets exhibiting tumbling (upper plot) and (lower plot) chaotic motions. (b) The complexity-entropy plane expresses the values of statistical complexity ($C$) versus permutation entropy ($H$) for each time series of the observable area of circular paper sheets. Green markers indicate motions classified as tumbling, while the purple markers represent chaotic behavior, both identified by a panel of human experts. (c) Automatic classification of the motion types into tumbling (green markers) and chaotic (light purple markers) obtained by applying the $k$-means clustering algorithm (with $k=2$ groups) to the same values of $H$ and $C$ shown in panel (b). The inset shows the silhouette score for different values of $k$ (number of groups), where we observe that $k=2$ yields the best cohesion and separation for our data. In panels (b) and (c), the continuous curves indicate the minimum and maximum complexity values for a given value of entropy, while the dashed lines represent the decision boundary obtained from the $k$-means algorithm. Supplementary Figures~1 and~2 show visualizations similar to panels (b) and (c) for the hexagon, square and cross shapes. (d) Classification agreement between our approach and the ``ground truth'' provided by the human experts for the four paper shapes. 
}
\label{fig:1}
\end{figure*}

We have focused our investigations on distinguishing between chaotic and tumbling behaviors as the steady motion is easily distinguishable from the two more complex behaviors~\cite{howison2020large-scale}. Furthermore, we have used the observable area time series $\{x_t\}_{t=1,\dots,N}$ ($N$ is the time series length) from one of the cameras as the only input used to classify the motion type as chaotic or tumbling. Figure~\ref{fig:1}(a) shows typical time series for each of these two free-falling behaviors. Values of $x_t\approx1$ indicate that the paper sheet is parallel to the camera viewing area, while for $x_t\approx0$ the paper is perpendicular to this area. These time series were recorded at a sample frequency of 98Hz (interval between consecutive observations is $\approx$0.01s). Table~\ref{tab:1} shows the total number of time series used in our work and the percentage of them exhibiting tumbling and chaotic behaviors for each paper shape.

\begin{table}[!t]
    \centering
    \caption{The total number of time series used in our work and the percentage of them exhibiting tumbling and chaotic behaviors for each paper shape.}
    \label{tab:1}\vspace*{0.5em}
    \begin{tabular}{lccc}
    \hline
    Shape   & Samples & Tumbling & Chaotic \\ \hline 
    Circle  &     170 &     49\% &    51\% \\ [.25em]
    Hexagon &     120 &     39\% &    61\% \\ [.25em]
    Square  &      92 &     30\% &    70\% \\ [.25em]
    Cross   &      59 &     31\% &    69\% \\ \hline
    \end{tabular}
\end{table}

\section{Methods and Results}\label{sec:methods}

We start by briefly revisiting the complexity-entropy plane approach~\cite{rosso2007distinguishing}. This method is a two-dimensional representation space comprising the values of permutation entropy~\cite{bandt2002permutation} and statistical complexity~\cite{lopezruiz1995statistical, rosso2007distinguishing}. These two measures are calculated using a probability distribution related to ordinal patterns (or permutation symbols) obtained directly from the time series. This process is usually known as the Bandt and Pompe symbolization process~\cite{bandt2002permutation}. 

To better describe this approach, let us consider again the time series $\{x_t\}_{t = 1, \dots, N}$ with $N$ elements representing the observable area of the motion of a given paper shape. First, we segment this time series using a sliding window of size $d$ (the so-called embedding dimension) that moves one element at a time. This time series segments can be written as 
\begin{equation}\label{eq:1dpartition}
    w_p =  (x_{p}, x_{p + 1}, x_{p + 2}, \dots, x_{p + (d - 2)}, x_{p + (d - 1)})\,,
\end{equation}
where $p = 1, \dots, n$ represents the partition index and $n = N - (d - 1)$ is the total number of partitions obtained after the sliding window reaches its last position. For each partition, we evaluate the permutation $\pi_p = (r_0, r_1, \dots, r_{d-1})$ of the index numbers $(0, 1, \dots, d - 1)$ that orders the elements of $w_p$, that is, the permutation of index numbers defined by the inequality $x_{p + r_0} \leq x_{p + r_1} \leq \dots \leq x_{p + r_{d-1}}$. If a partition contains repeated values, $x_{p+r_{j-1}} = x_{p+r_j}$, we maintain the occurrence order by setting $r_{j-1} < r_{j}$ for $j=1,\dots,d-1$~\cite{cao2004detecting}. To illustrate this procedure, suppose we have the time series $x_t = (3,2,2,7,9,5)$ and set the sliding window size $d = 3$. The first partition is $w_1 = (3,2,2)$, and by sorting its elements, we find $2 \leq 2 < 3$ or $x_{1+1} \leq x_{1+2} < x_{1+0}$. Therefore, the ordinal pattern associated with $w_1$ is $\pi_1 = (1,2,0)$.

After evaluating the permutation for each partition, we obtain a sequence of permutations (ordinal patterns) $\{\pi_p\}_{p = 1, \dots, n}$. Using this sequence, we estimate the ordinal probability distribution $P = \{\rho_j(\Pi_j)\}_{j = 1, \dots, d!}$ of finding each $\Pi_j$ of the $d!$ possible ordinal patterns by calculating their relative frequencies of occurrence, that is, 
\begin{equation}\label{eq:permutation_probability}
    \rho_j(\Pi_j) = \frac{\text{number of partitions of type} \ \Pi_j \ \text{in} \ 
    \{\pi_p\}}{n}\,.
\end{equation}
The permutation entropy~\cite{bandt2002permutation} is simply the Shannon entropy~\cite{shannon1948mathematical} of the ordinal probability distribution $P$, 
\begin{equation}\label{eq:permutation_entropy}
    S(P) = -\sum_{j = 1}^{d!} \rho_j(\Pi_{j})\log \rho_j(\Pi_{j})\,.
\end{equation}
Because the permutation entropy reaches its maximum value ($S_{\rm max} = \log{d!}$) for a uniform ordinal distribution ($U=\{1/d!\}_{j = 1, \dots, d!}$), we can further define its normalized version as
\begin{equation}\label{eq:normalized_pe}
    H(P) = \frac{S(P)}{\log{d!}}\,.
\end{equation}
The values of $H$ measure the degree of randomness in the time series such that $H\approx0$ for a completely regular series while $H\approx0$ indicates a completely random series. In its turn, the statistical complexity $C$~\cite{lopezruiz1995statistical, rosso2007distinguishing} is defined as a product between the normalized permutation entropy $H$ and the normalized Jensen-Shannon divergence $D(P,U)$ between the system's ordinal distribution ($P$) and the uniform distribution ($U$), that is,
\begin{equation}\label{eq:statistical_complexity}
    C(P) = \frac{D(P,U)H(P)}{D_{\rm max}}\,,
\end{equation}
where
\begin{equation}\label{eq:js_divergence}
    D(P,U) = S[(P + U)/2] - \dfrac{1}{2}S(P) - \dfrac{1}{2}S(U),
\end{equation}
is the Jensen-Shannon divergence and
\begin{equation*} 
    D_{\rm max} = -\dfrac{1}{2}\left(\frac{d!+1}{d!}\log(d!+1)-2\log(2d!)+\log{d!}\right)
\end{equation*}
is a normalization constant corresponding to the maximum value of $D(P,U)$ that occurs when one component of $P$ is equal to one and all others are zero. Differently from permutation entropy, the statistical complexity $C$ is close to zero in both extremes of order and disorder, whereas $C>0$ indicates larger structural complexity in the ordering dynamics of the time series. The combined use of the values of $C$ versus $H$ defines a representation space that is called the complexity–entropy plane~\cite{rosso2007distinguishing}. It is always important to remark that because $H$ and $C$ depend on different sums of the ordinal distribution (see Eq.~\ref{eq:js_divergence}), there is no reason for assuming a univocal relationship between these two quantities. Indeed, for a given value of entropy $H$ there exists a range of possible values for complexity $C$~\cite{lamberti2004intensive, martin2006generalized, rosso2007distinguishing}, allowing the complexity to provide additional information about the system dynamics.

The only parameter of this approach is the embedding dimension $d$; however, its choice is not entirely arbitrary as one needs to satisfy the condition $d!\ll N$ to have a reliable estimate of the ordinal distribution~\cite{rosso2007distinguishing}. In our case, the average time series length is approximately $110$ elements, and because of that, we have fixed $d=3$ in all our results. For the numerical implementation of the complexity-entropy plane, we rely on the Python module \texttt{ordpy}~\cite{pessa2021ordpy}.

We have thus evaluated the values of permutation entropy and statistical complexity for all time series of the observable area after grouping the data by paper shape. Figure~\ref{fig:1}(b) depicts the complexity-entropy plane for circular paper shapes, where motions classified by experts as tumbling and chaotic are colored in green and light purple, respectively. This figure also shows the curves of minimum (lower continuous line) and maximum (upper continuous line) complexity values as a function of the entropy values. We observe that time series related to tumbling motions are located in a region of the plane characterized by higher entropy and lower complexity values than time series from chaotic motions. Moreover, chaotic time series are located near the maximum values of complexity, while tumbling motion is almost equidistant to minimum and maximum complexity values. These results somehow resemble the distinction between chaotic maps and simulated stochastic processes reported in the seminal work of Rosso~\textit{et al.}~\cite{rosso2007distinguishing}. In their work, Rosso~\textit{et al.}~\cite{rosso2007distinguishing} found that time series from several chaotic maps are located near the maximum complexity values while series from stochastic processes show intermediate values of complexity. Supplementary Figure~1 shows that these patterns hold not only for circular papers but also for hexagonal, square, and cross shapes.

These results demonstrate that the observable area of free-falling papers has a distinct average degree of entropy and complexity, which may allow us to automatically cluster these types of motions using unsupervised machine learning methods. To test this possibility, we have applied the $k$-means clustering algorithm~\cite{james2014introduction, gueron2017handson} to the values of permutation entropy and statistical complexity related to all time series (as implemented in the Python module \texttt{scikit-learn}~\cite{pedregosa2011scikitlearn}). This method splits an unlabeled dataset into $k$ clusters by optimizing the cluster centers to make each observation closer to its own cluster center than to all other cluster centers. Figure~\ref{fig:1}(c) shows the partition of the points in the complexity-entropy plane for circular shapes into two clusters ($k=2$), where the dashed line represents the decision boundary separating the two groups [this line is also shown in Fig.~\ref{fig:1}(b)]. Furthermore, since the number of clusters is predefined in the $k$-means algorithm, we have investigated whether grouping the dataset into two clusters represents an optimal partitioning scheme. To do so, we estimate the silhouette score~\cite{rousseeuw1986silhouettes, gueron2017handson} of partitions obtained from the $k$-means algorithm with different values of $k$. Ranging from $-1$ to $1$, the silhouette score quantifies cohesion and separation of grouped data, such that the higher its value the better the clustering configuration. Results in the inset of Fig.~\ref{fig:1}(c) show that $k=2$ yields the highest silhouette score and so a partition into two groups can be considered optimal in terms of this metric. As shown in Supplementary Figure~2, we obtain the same conclusion for all paper shapes and also when using the ``elbow method'' with the cluster inertia (within-cluster sum-of-squares)~\cite{gueron2017handson}.

We can further quantify the performance of our clustering approach by comparing the groups obtained from the complexity-entropy plane with the classification made by experts~\cite{howison2020large-scale}. To do this, we evaluate the classification agreement between our automated clusters and the experts' classifications. Figure~\ref{fig:1}(d) shows that this agreement ranges from 87\% for the square-shaped papers to 69\% for the cross-shaped ones. These agreement scores are significantly higher than those obtained by applying our approach to shuffled versions of the observable area time series (average agreement close to random guesses), which confirms that the ordering dynamics carry information for distinguishing between the two types of motion. Also, they outperform the simple strategy of labeling all time series with the most common behavior (Table~\ref{tab:1}). Our agreement is also quite similar to the one reported by Howison \textit{et al.}~\cite{howison2020large-scale} using a diagram formed by the values of the non-dimensional moment of inertia versus the Reynolds number. However, it is worth noticing that the evaluation of these two dimensionless quantities depends on extracting the three-dimensional trajectory of the paper shapes~\cite{howison2020large-scale}, while entropy and complexity values are estimated directly from the observable area of the paper --- a one-dimensional series much more easily obtained from experiments than the complete trajectory. Finally, and as also argued by Howison \textit{et al.}~\cite{howison2020large-scale}, it is worth remarking that the visual distinction between chaotic and tumbling motions is sometimes ambiguous such that human experts cannot agree on what behavior they are seeing. For instance, it seems to us that the trajectory classified as tumbling by the experts but showing the highest complexity value of its class in Figure~\ref{fig:1}(b) has patterns much closer to other chaotic than tumbling behaviors (a hypothesis that is reinforced after watching the video of this particular falling paper). Thus, our less subjective classification approach based on the complexity-entropy plane can also reduce the ambiguity of human classifications.

\section{Conclusions}\label{sec:conclusions}

We have presented an approach capable of accurately and automatically distinguishing between chaotic and tumbling motions of pieces of paper falling through the air. Our method is based on estimating the complexity-entropy plane for time series related to the observable area of falling papers, and so it naturally inherits all advantages of ordinal methods (simplicity, low computational cost, noise robustness, and invariance against monotone scaling of the data). Furthermore, in comparison with other approaches based on physical quantities (such as the non-dimensional moment of inertia and the Reynolds number), this information-theory framework has the advantage of depending only on one-dimensional time series and not requiring the complete reconstruction of the three-dimensional motion of falling papers. Our work also belongs to a long and successful tradition of applying informational measures to study complex time series~\cite{costa2005multiscale, rostaghi2016dispersion, guntu2020wavelet, guntu2020spatiotemporal}. We further believe that our work may trigger other investigations focused on using the extensive flora of ordinal methods and that there is certainly room for improving the accuracy in distinguishing and perhaps even identifying types of motions associated with the falling-paper problem. Finally, we hope that our work contributes to a better understanding of falling papers’ behaviors and motivates further empirical investigations about nonlinear time series, a still somewhat rare to find application in the context of ordinal methods.

\acknowledgments
We acknowledge the support of the Coordena\c{c}\~ao de Aperfei\c{c}oamento de Pessoal de N\'ivel Superior (CAPES), the Conselho Nacional de Desenvolvimento Cient\'ifico e Tecnol\'ogico (CNPq -- Grant 303533/2021-8), and the Slovenian Research Agency (Grants J1-2457 and P1-0403).


\clearpage
\includepdf[pages=1-2,pagecommand={\thispagestyle{empty}}]{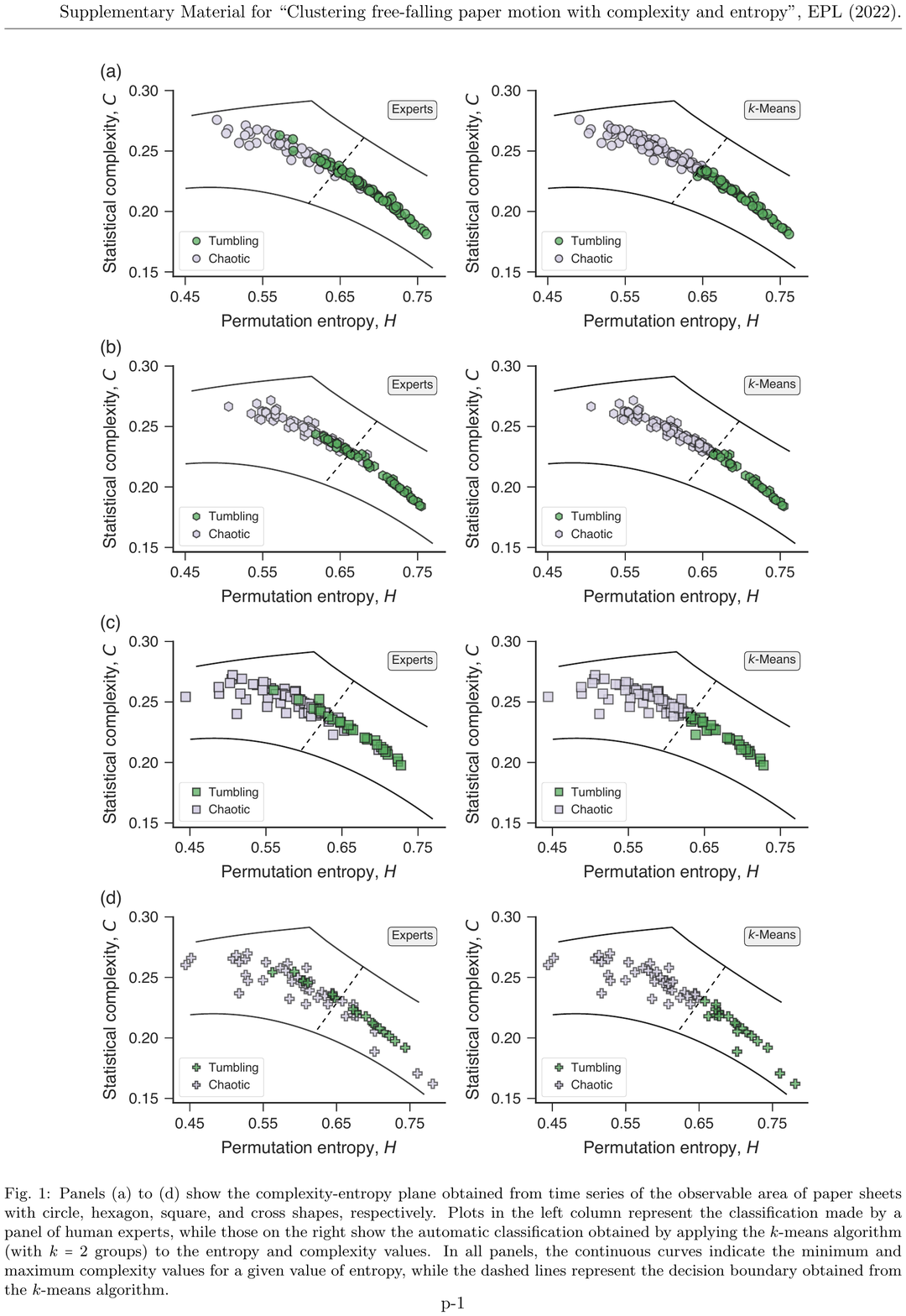}

\end{document}